\documentclass[pre,showpacs,floatfix]{revtex4}
\usepackage{graphicx,bm}
\begin{document}

\title{Correlation Function Bootstrapping in Quantum Chaotic Systems}
\author{L. Kaplan}
\affiliation{Department of Physics, Tulane University, New Orleans, Louisiana
70118 USA}
\date{March 24, 2005}

\begin{abstract}
We discuss a general and efficient approach for ``bootstrapping" short-time
correlation data in chaotic or complex quantum systems to obtain information
about long-time dynamics and stationary properties, such as the local density
of states.  When the short-time data is sufficient to identify an individual
quantum system, we obtain a systematic approximation for the spectrum and wave
functions.  Otherwise, we obtain statistical properties, including wave
function intensity distributions, for an ensemble of all quantum systems
sharing the given short-time correlations.  The results are valid for open or
closed systems, and are stable under perturbation of the short-time input data.
Numerical examples include quantum maps and two-dimensional anharmonic
oscillators.
\end{abstract}


\maketitle

\section{Introduction}
\label{secintro}

When a quantum system is known to have a chaotic classical limit, the simplest
description of its eigenvalues, eigenstates, and dynamics is given by the
universal predictions of random matrix theory (RMT)~\cite{rmt} and the closely
related random wave hypothesis~\cite{berryrw}.  Recently, however, there has
been increased interest in understanding the deviations from RMT that are quite
sizable in many systems of interest and are often due to short-time dynamical
effects.  RMT assumes implicitly that under time evolution, any initial state
immediately spreads randomly over the entire available Hilbert space; any
realistic chaotic dynamics, however, maintains at short times information about
the initial state and only after some finite mixing time $T_{\rm mix}$ does the
dynamics become truly random.

For the purpose of describing spectral statistics, such as the distribution of
level spacings, the above consideration may easily be taken into account by
noting that RMT predictions are valid only inside energy windows of size less
than a ballistic Thouless energy $E_c \sim \hbar/T_{\rm mix}$.  The situation
with wave functions is not so simple, as short-time dynamical effects can lead
to large deviations from RMT not only for energy-averaged quasimodes but also
for individual eigenstates.

Any short unstable periodic orbit is one obvious example of a non-random
dynamical feature that is known to cause significant deviations from RMT
statistics in the eigenstates.  It has been shown that the ``scarring" of
individual wave functions by a typical periodic orbit is an $O(\hbar^0)$ effect
that persists in the semiclassical limit,  as measured by the distribution of
wave function intensities on the periodic orbit in Husimi phase
space~\cite{scarhusimi}.  Using a linearization of the dynamics around the
specific orbit, the distribution of quantum intensities on or near the orbit
may be obtained semiclassically as a function of the classical monodromy
matrix, and various moments of the distribution, such as the inverse
participation ratio, may be expressed analytically~\cite{scarmom}.  The
scarring effect has been studied experimentally and numerically in a wide
variety of systems~\cite{scareffect}, and may have consequences for the
conductance through a resonant tunneling diode or a ballistic quantum
dot~\cite{scardiode,scarcond}.

The imprint of short-time dynamics on eigenstate structure is not limited to
situations where the short-time dynamics is related to classical unstable
periodic orbits.  For example, scar-like resonances related to diffractive
trajectories have been observed in a 2DEG conductance experiment~\cite{diffr}.
The approach of using short-time behavior to understand eigenstate structure
and statistics has been used successfully in many situations where a proper
classical limit does not exist, such as two-body random interactions in a
many-fermion system~\cite{papenbrock} or dynamics on a quantum graph or
lattice~\cite{graphs}, as well as in pseudo-integrable systems where the
Lyapunov exponent vanishes~\cite{superscar}.  Furthermore, short-time dynamical
information necessarily implies deviations from RMT not only for individual
wave function intensities, but also for spatial or phase space
correlations~\cite{schanz}.

Our aim here is to discuss a systematic, general, and efficient framework for
studying the constraints that short-time correlations impose on the eigenstate
structure, regardless of whether such short-time correlations can be computed
semiclassically.  We allow ourselves to focus on one or an arbitrary number of
initial wave packets, and the calculations may be performed equally well in
closed or open systems, as there is no assumption of unitarity in the dynamics.
We explicitly allow for the presence of errors in the short-time correlations,
and demonstrate the stability of the results with respect to such errors.  

In the context of extracting stationary properties from a time-domain
correlation function, we mention the important work that has been done by
Mandelshtam and coworkers using the ``filter diagonalization"
method~\cite{mandelshtam}.  In that approach, one typically begins with a
single wave packet, and spectral information can in principle be computed {\it
exactly} when the correlation function is known for at least $N$ times, where
$N$ is the Hilbert space dimension.  In the bootstrapping approach, we use
multiple initial wave packets, and we do not assume exact
finite-dimensionality of the Hilbert space.  Thus our goal is not an exact
solution of the spectral analysis problem, but rather an increasingly good
approximation as the amount of input data increases.  Of course, exact
solutions are not possible in any case in the presence of noise or numerical
instability, so in practice regularization must be performed, yielding
comparable results for the two approaches.  One advantage of the bootstrapping
approach is that the linear algebra involved requires $M \times M$ matrices
only, where $M$ is the number of initial wave packets, allowing the problem to
scale very well computationally for large system sizes and long times.

The paper is organized as follows.  In Sec.~\ref{secboot}, we define the
quantities of interest in the time and energy domain, and obtain general
expressions for the bootstrapped correlation function and spectrum.
Sec.~\ref{secmodels} serves to define two sets of numerical models, which may
be used to illustrate the general formalism.  Next, in Sec.~\ref{secconv}, we
discuss convergence properties of the bootstrapping approximation, including a
treatment of stability in the presence of noise.  In Sec.~\ref{secstat}, we
examine how bootstrapping may be used to compute statistical quantities of
interest, including wave function intensity distributions and intensity
correlations, using a very small amount of time-domain data as input.  Finally,
the key conclusions are briefly summarized in Sec.~\ref{secsum}.

\section{Bootstrapping for Correlation Functions and Spectrum}
\label{secboot}

We begin by considering a set of $M$ wave packets $\phi_i$, for simplicity
taking the wave packets to be orthonormal (but not a complete set).  In
practice, the choice of $\phi_i$ will be dictated by the physics of interest.
For example, the $\phi_i$ may be chosen as position eigenstates if we are
interested in position-space wave function structure, or momentum eigenstates
for scattering behavior, or Gaussian wave packets for analyzing the effects of
classical periodic orbits.  In a many-body problem, the $\phi_i$ may usefully
be taken as the non-interacting product states.

The quantity of interest in the time domain is the correlation function
\begin{equation}
\label{correlfcn}
C_{ij}(t)=\langle \phi_i |e^{-i\hat Ht/\hbar} |\phi_j \rangle \,,
\end{equation}
whose diagonal elements $C_{ii}(t)$ constitute the autocorrelation functions
for the individual wave packets.  Knowledge of the exact correlation function
for all discrete times $t=mT_0$ leads by Laplace transform to the discrete-time
Green's function
\begin{eqnarray} G_{ij}(E)&=&
 (i \hbar)^{-1} T_0 \sum_{m=0}^\infty \left [e^{iEmT_0/\hbar} 
-{1 \over 2} \delta_{m0}\right]
C_{ij}(mT_0)
 = (i\hbar)^{-1} T_0
\langle \phi_i |{ 1\over 1-e^{i(E - \hat H)T_0/\hbar}} -{1 \over 2}| \phi_j
\rangle \\ & \approx &(i \hbar)^{-1} \int_0^\infty dt\; e^{iEt/\hbar}
C_{ij}(t)=
\langle \phi_i |{ 1\over E - \hat H + i \epsilon} | \phi_j \rangle \,,
\label{greenconti}
\end{eqnarray}
where the continuous-time limit of Eq.~(\ref{greenconti}) is obtained for $T_0
\ll \hbar/\delta E$, and $\delta E$ is a typical energy spread in the wave
packets.  

We will see that it proves useful to decompose the return amplitude of
Eq.~(\ref{correlfcn}) at time $t$ into a special ``new" component that is
returning for the first time to the subspace spanned by the $M$ wave packets
$\phi_i$ and the remainder due to propagation that has revisited this subspace
at least once at time steps in between $0$ and $t$.  In spirit, this is
reminiscent of the $T-$matrix approach of Bogomolny~\cite{tmatrix}, where an
integral kernel is defined in terms of all classical trajectories that start on
a given Poincar\'e surface of section and return to the surface of section
without intersecting it at intermediate times.  The $T-$matrix is defined
directly in the energy domain whereas we begin our analysis in the time domain
and transform to the energy domain later on.  The decomposition used here
also resembles somewhat the one used in the quantitative analysis of periodic
orbit scarring~\cite{scarhusimi}, where it is helpful to separate terms in the
return amplitude that are associated with paths staying on the periodic orbit
from terms associated with homoclinic paths that leave the orbit once and first
return at some later time.  Of course, in the case we consider here the
$M-$dimensional subspace spanned by the wave packets $\phi_i$ will not in
general have any connection with a particular classical structure such as a
periodic orbit or surface of section.  Instead the choice of $\phi_i$ is
governed either by our exact or approximate knowledge of the correlation
function for those initial and final states or by an interest in extracting
wave function structure information in a specific basis or phase space region.
Furthermore, no semiclassical approximation is implicit in the method we
develop here, although  we will see below that the approach can be extended to
situations where the correlation function information used as input is only
approximate, as would be the case for example when a semiclassical propagator
is used.

Formally, the new recurrences $B_{ij}(m)$ at time $t=mT_0$ may be defined as
\begin{equation}
B_{ij}(m)= \langle \phi_i |e^{-i\hat H T_0/\hbar}
\left ((1-\hat P)e^{-i\hat H T_0/\hbar}\right )^{m-1} |\phi_j\rangle \,,
\end{equation}
where
\begin{equation}
\hat P= \sum_{k=1}^M |\phi_k \rangle \langle \phi_k|
\end{equation}
is the projection onto the subspace of interest.  More explicitly, these
new recurrences may be computed from the full $C_{ij}$ amplitudes as
\begin{equation}
\label{bijdef}
B_{ij}(m) = \left \{ \begin{array}{ll}
C_{ij}(T_0) & m=1 \\
C_{ij}(mT_0) - \sum_{p=1}^{m-1} \sum_{k=1}^M
B_{ik}(p)C_{kj}((m-p)T_0) \;\;& m\ge 2 
\end{array} \right. \,.
\end{equation}
The full correlation function is then given by a convolution,
\begin{equation}
\label{cindex}
C_{ij}(mT_0)=\sum_{p=1}^m \sum_{k=1}^M B_{ik}(p) C_{kj}((m-p)T_0)
\end{equation}
or, in matrix notation,
\begin{eqnarray}
\label{csum}
{\mathbf C}(mT_0)&=&\sum_{p=1}^m {\mathbf B}(p) {\mathbf C}((m-p)T_0)
\nonumber \\ &=& {\mathbf B}(m) + \sum_{p=1}^{m-1}{\mathbf B}(p){\mathbf
B}(m-p)+\sum_{p=1}^{m-2}\sum_{p'=1}^{m-p-1}
{\mathbf B}(p){\mathbf B}(p'){\mathbf B}(m-p-p') + \cdots \,,
\end{eqnarray}
where ${\mathbf C}(0)$ is always the identity matrix.  The matrix ${\mathbf
B}(m)$ records amplitude that at the $m-$th step returns for the first time to
the subspace spanned by the $\phi_i$, while terms in Eq.~(\ref{csum}) involving
a product of $n$ $\mathbf B$-matrices correspond to processes where amplitude
leaves and returns $n$ times to the same subspace over $m$ steps.  In the
energy domain,
\begin{equation}
\label{gtildeb}
G_{ij}(E)=(i \hbar)^{-1} T_0\langle \phi_i| {1 \over 1- {\mathbf {\tilde B}}(E)}
-{1 \over 2}| \phi_j \rangle \,,
\end{equation}
where
\begin{equation}
\label{tildeb}
{\mathbf {\tilde B}}(E) = \sum_{m=1}^\infty e^{i m T_0 E/\hbar} {\mathbf B}(m)
\,.
\end{equation}

We are however interested in the information that can be extracted from
knowledge of the correlation function at a finite set of times $t$ only, say
$t=m T_0$ for $m = 1 \cdots L$, possibly in the presence of noise.  If we
assume ${\mathbf C}(mT_0)$ is known only for times $t \le T_{\rm max}=LT_0$,
i.e., for $1 \le m \le L$, then we may compute the new recurrences ${\mathbf
B}(mT_0)$ only for $1 \le m \le L$ using Eq.~(\ref{bijdef}).  It is convenient
to define
\begin{equation}
{\mathbf B}_{L,\tau}(m)= \left \{ \begin{array}{ll} {\mathbf
B}(m)e^{-mT_0/\tau}\;\; & 1 \le m \le L \\ 0 & {\rm otherwise} \end{array}
\right . \,.
\label{bltau}
\end{equation}
The cutoff time $\tau$, which can be much larger than the bootstrap time
$T_{\rm max}=LT_0$, serves as a smoothing scale in the energy domain, and its
significance will be discussed in Sec.~\ref{secconv} below.  Given just the
matrices ${\mathbf B}_{L,\tau}(m)$, we may compute a ``bootstrapped"
approximation to the full correlation function at all times:
\begin{eqnarray}
{\mathbf C}_{L,\tau}(mT_0)&=&\sum_{p=1}^m {\mathbf B}_{L,\tau}(p) {\mathbf
C}_{L,\tau}((m-p)T_0)  \nonumber \\ &=& {\mathbf B}_{L,\tau}(m) +
\sum_{p=1}^{m-1}{\mathbf B}_{L,\tau}(p){\mathbf
B}_{L,\tau}(m-p)+\sum_{p=1}^{m-2}\sum_{p'=1}^{m-p-1} {\mathbf
B}_{L,\tau}(p){\mathbf B}_{L,\tau}(p'){\mathbf B}_{L,\tau}(m-p-p') + \cdots\,,
\label{csuml}
\end{eqnarray}
having the property ${\mathbf C}_{L,\tau}(mT_0)={\mathbf
C}(mT_0)e^{-mT_0/\tau}$ for $m \le L$.  In the energy domain, ${\mathbf {\tilde
B}}_{L,\tau}(E)$ may be defined as a Laplace transform of ${\mathbf
B}_{L,\tau}(m)$, in complete analogy with Eq.~(\ref{tildeb}) above.

Finally, one often encounters a ``noisy" situation where even the short-time
dynamics is only approximately known.  For example, we may be interested in
building up the full dynamics using only {\it semiclassical} expressions for
the propagator at short times.  We then have knowledge of
\begin{equation}
\label{defeps}
{\mathbf C}_\epsilon(mT_0) = {\mathbf C}(mT_0)+ \epsilon {\mathbf D}(mT_0)
\end{equation}
for $1 \le m \le L$, where ${\mathbf D}(mT_0)$ are quasi-random, uncorrelated
error matrices and $\epsilon$ characterizes the relative size of the error.
Given this input data we may calculate approximate ``new" recurrences ${\mathbf
B}_{L,\tau,\epsilon}(mT_0)$ by extending the exact formula of
Eq.~(\ref{bijdef}),
\begin{equation}
{\mathbf B}_{L,\tau,\epsilon}(m) = \left \{ \begin{array}{ll}
{\mathbf C}_\epsilon(T_0)e^{-T_0/\tau} & m=1 \\
{\mathbf C}_\epsilon(mT_0)e^{-mT_0/\tau} - \sum_{p=1}^{m-1}
{\mathbf B}_{L,\tau,\epsilon}(p){\mathbf
C}_\epsilon((m-p)T_0)e^{-(m-p)T_0/\tau} \;\;& 2 \le m \le L \\ 0 & {\rm
otherwise}
\end{array} \right. \,.
\end{equation}
The ``bootstrapped" long-time evolution ${\mathbf C}_{L,\tau,\epsilon}$ is
given by iterating these approximately known short-time ``new" recurrences
analogously to Eq.~(\ref{csuml}),
\begin{eqnarray}
{\mathbf C}_{L,\tau,\epsilon}(mT_0)&=&\sum_{p=1}^m {\mathbf
B}_{L,\tau,\epsilon}(p) {\mathbf C}_{L,\tau,\epsilon}((m-p)T_0)  \nonumber \\
&=& {\mathbf B}_{L,\tau,\epsilon}(m) + \sum_{p=1}^{m-1}{\mathbf
B}_{L,\tau,\epsilon}(p){\mathbf
B}_{L,\tau,\epsilon}(m-p)+\sum_{p=1}^{m-2}\sum_{p'=1}^{m-p-1} {\mathbf
B}_{L,\tau,\epsilon}(p){\mathbf B}_{L,\tau,\epsilon}(p'){\mathbf
B}_{L,\tau,\epsilon}(m-p-p') + \cdots\,.
\end{eqnarray}

Again, by construction the bootstrapping procedure simply reproduces the noisy
input data for times $t$ below the bootstrap time $T_{\rm max}$, i.e.,
${\mathbf C}_{L,\tau,\epsilon}= {\mathbf C}_\epsilon e^{-mT_0/\tau}$ for $m \le
L$.  However, bootstrapping allows us also to learn something about longer
times $t > T_{\rm max}$ using the short-time correlation function.

\section{Numerical Models}
\label{secmodels}

\subsection{Quantum Maps}
\label{secmaps}

Classical and quantum chaotic maps in one dimension are often used as the
simplest examples for illustrating general chaotic behavior, and share many
scaling and other physical properties of two-dimensional Hamiltonian
dynamics~\cite{maps}.  Discrete-time maps may be thought of as arising from a
continuous-time Hamiltonian dynamics either via a Poincar\'e surface of section
or by stroboscopically viewing motion in a periodically driven Hamiltonian.  In
the latter case, we may consider
\begin{equation}
H(q,p,t) = { 1 \over T_{\rm kick}}
T(p) +V(q) \sum_{j=-\infty}^\infty \delta(t-jT_{\rm kick}) \,,
\end{equation}
which yields
\begin{eqnarray}
p_{j+1} &=& p_j-V'(q_{j}) \nonumber
 \\
q_{j+1}&=&q_j+T'(p_{j+1})
\,,
\label{quanmap}
\end{eqnarray}
when the position $q_j$ and momentum $p_j$ are recorded just before kick $j$.
The corresponding quantum evolution over one time step is given by
\begin{equation}
\label{quantmap}
\hat U = e^{-i T(\hat p)/\hbar} e^{-i V(\hat q)/\hbar} \,.
\end{equation}

As a specific example, we may take $T(p)= {1 \over 2} w_p p^2+K_p (\sin{p}- {1
\over 2} \sin{2p})$ and $V(q)=-{ 1\over 2} w_q q^2-K_q (\sin{ q}-{1\over 2}
\sin{2q})$, for a toral phase space $(q,p) \in [-\pi0,\pi)\times [-\pi,\pi)$.  With
integer values of $w_q$ and $w_p$, this is a perturbed cat map~\cite{pertcat}:
\begin{eqnarray}
p_{j+1} &=& p_j +w_q q_j +K_q (\cos{q_j}-\cos{2q_j}) \;\; {\rm mod} \;\; 2 \pi \nonumber \\
q_{j+1} &=& q_j +w_p p_j +K_p (\cos{p_j}-\cos{2p_j}) \;\; {\rm mod} \;\; 2 \pi \,,
\label{pertmap}
\end{eqnarray}
where nonzero values of $K_q$, $K_p$ are needed to break the symmetries and
ensure nonlinearity of the dynamics.  One easily checks that the dynamics is
completely chaotic for sufficiently small $K_{q,p}$.

For this compact classical phase space, the quantum evolution of
Eq.~(\ref{quantmap}) acts on a Hilbert space of dimension $N=2 \pi/\hbar$, the
mean energy level spacing is $\Delta= 2 \pi \hbar/NT_{\rm kick}=\hbar^2/T_{\rm
kick}$, and the Heisenberg time at which levels are resolved is $T_H=N T_{\rm
kick}$.  Since the map dynamics is already discretized, it is natural to use
the period $T_{\rm kick}$ as the time step $T_0$ in the bootstrapping
calculation.  Without loss of generality, we may choose units where $T_0
=T_{\rm kick}=1$.

\subsection{Two-Dimensional Wells}

As our model of a non-integrable system with a time-independent Hamiltonian, we
use the Barbanis Hamiltonian~\cite{barbanis}, which describes a two-dimensional
anharmonic oscillator:
\begin{equation}
H(x',y',{p'}_x,{p'}_y)={{p'}_x^2 \over 2m}+ {{p'}_y^2 \over 2m}+
{1 \over 2} m \omega_x^2 x'^2 + {1 \over 2} m \omega_y^2 y'^2 +
\lambda x'{y'}^2 \,.
\end{equation}
After a canonical transformation and an overall rescaling of the energy, the
Barbanis Hamiltonian may be re-written as
\begin{equation}
\label{barbeqn}
H(x,y,p_x,p_y)={p_x^2 \over 2}+ {p_y^2 \over 2}+
{1 \over 2} x^2 + {a \over 2} y^2 +
{a \over 2} xy^2 \,,
\end{equation}
where $a$ is a dimensionless parameter characterizing the shape of the well.
In these dimensionless coordinates, the metastable well has its minimum at
$x=y=0$ and extends from $x=-1$ to $x=1$ along the $y=0$ symmetry axis; the
barrier height is $E_{\rm max}=1/2$.  Upon quantization, one additional
parameter besides $a$ is introduced, namely an effective $\hbar$ or
equivalently the number of quantum levels below $E_{\rm max}$.

As compared with the simple quantum map model presented above, analysis of
bootstrapping in the Barbanis system requires consideration of the following
three circumstances, which are typical of many Hamiltonian systems: (i) time is
not naturally discrete and thus an explicit choice is needed for the
discretization time $T_0$, (ii) resonances in the metastable well have finite
intrinsic width, introducing a new long-time scale, and (iii) classical
dynamics in the well is mixed, with the phase space at energies of interest
shared by regular islands and a chaotic sea.  The implications of these three
circumstances will be discussed below, when numerical results for bootstrapping
in the Barbanis system are presented in Sec.~\ref{secenergydom}.

\section{Convergence Properties and Sensitivity to Error}
\label{secconv}

In this section, we examine how bootstrapping may be used when the given
information about the short-time correlation function is sufficient to compute
(approximately) the long-time dynamics and spectrum.  An alternative situation,
where the given information only restricts us to an ensemble of possible
long-time behaviors, and the objective is to obtain {\it statistical}
properties of the long-time dynamics or spectrum, is discussed in
Sec.~\ref{secstat}.

\subsection{Noise-Free Bootstrapping}
\label{secnoisefree}

We want to estimate the error made in using short-time information up to the
bootstrap time $T_{\rm max}=LT_0 $ to estimate long-time dynamics in a chaotic
system at times $t \gg T_{\rm max}$.  Let us first assume negligible noise by
setting $\epsilon=0$ in Eq.~(\ref{defeps}).  Clearly the error is then
associated with amplitude that starts in the subspace spanned by the $M$ wave
packets $\phi_i$ and is never captured by the short-time correlation function
because it does not return to the original subspace at any time during the
first $L$ steps of evolution.  In terms of the ${\mathbf B}-$matrices discussed
in the previous section, the total probability that does not return in time
$T_{\rm max}=L T_0$ is
\begin{equation}
\label{probnoreturn}
P(T_{\rm max})= 1- {1 \over M} \sum_{m=1}^L 
{\rm Tr} \; {\mathbf B}(m)^\dagger {\mathbf B}(m) \,.
\end{equation}
Mathematically, the probability $P(T_{\rm max})$ is clearly related to the
probability of remaining for at least time $T_{\rm max}$ in a system with $M$
maximally coupled open decay channels.  When the dynamics is chaotic, this
probability can be represented analytically as an integral in the context of
random matrix theory~\cite{savin}; for our purposes it is sufficient to note
that
\begin{equation}
P(T_{\rm max}) = \left \{ \begin{array}{ll}
e^{-MT_{\rm max}/T_H} & T_{\rm max} \ll T_H /\sqrt{M} \\
{1 \over M+1} \left({T_H / T_{\rm max}}\right ) ^{M+1}  & T_{\rm max} \gg T_H
\end{array} \right . \,,
\label{plost}
\end{equation}
where $T_H$ is the Heisenberg time, and the power-law long-time limit also
serves as an upper bound for $P(T_{\rm max})$.  Clearly, we require $T_{\rm
max} > T_H/M$ in order to recapture most of the initial amplitude, so that the
lost probability is small.  We emphasize that this estimate, based on random
matrix theory, may be used to obtain the correct scaling behavior of the lost
probability $P(T_{\rm max})$ with bootstrap time $T_{\rm max}$, even when the
prefactor in Eq.~(\ref{plost}) is invalid due to nonrandom short-time dynamical
effects.

\begin{figure}[ht]
\centerline{\includegraphics[width=4.5in,angle=270]{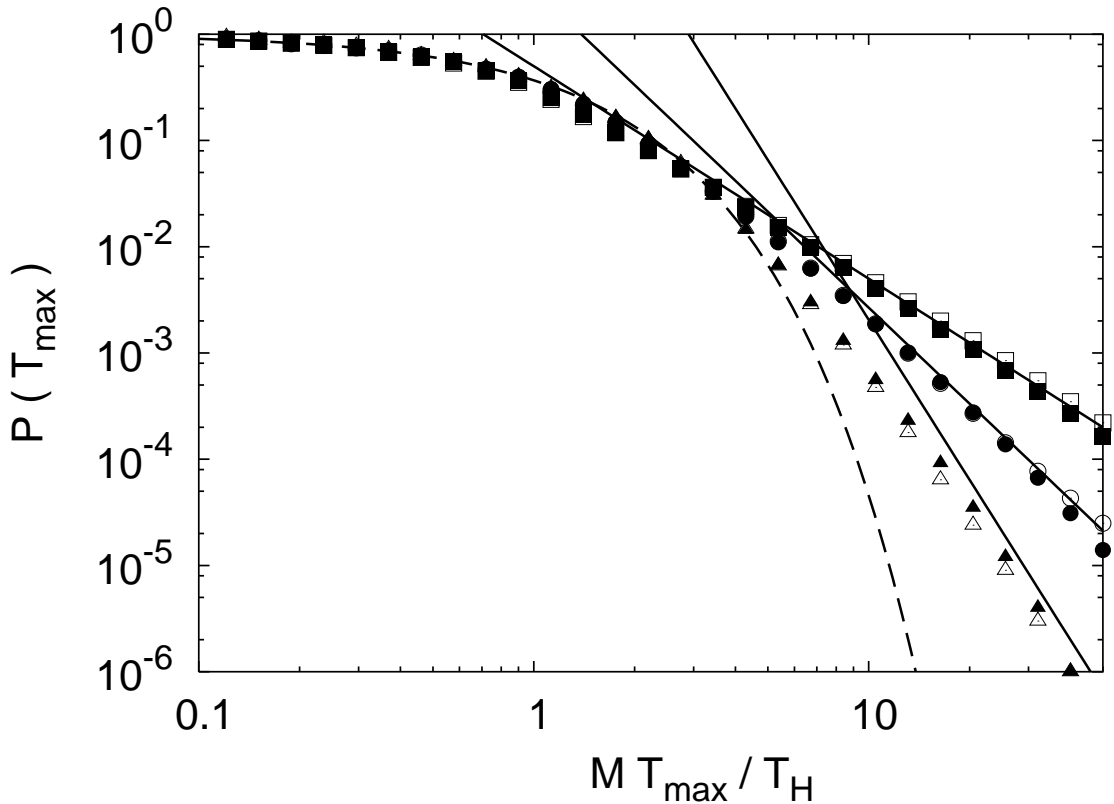}}
\vskip 0.2in
\caption{
The lost probability $P(T_{\rm max})$ associated with eliminating new
recurrences arriving after time $T_{\rm max}$ is plotted as a function of
$T_{\rm max}$ for the quantum map of Eq.~(\ref{pertmap}), with $w_q=w_p=1$ and
parameters $K_q$, $K_p$ uniformly distributed between $-1/2$ and $1/2$.
Squares, circles, and triangles represent data when the number of wave packets
$M$ is $M=1$, $2$, or $4$, respectively.  Open symbols are associated with
system size $N=64$ and closed symbols with system size $N=128$.  The dashed
curve is the small-$T_{\rm max}$ (classical) limit in Eq.~(\ref{plost}), while
the three solid lines represent the large-$T_{\rm max}$ power-law falloff for
$M=1$, $2$, and $4$.  All quantities shown in this and subsequent figures
are dimensionless.
}
\label{fig_vart}
\end{figure}

The behavior of the lost probability $P(T_{\rm max})$ for small and large
$T_{\rm max}$ is illustrated in Fig.~\ref{fig_vart}.  Here an average over
quantum maps given by Eq.~(\ref{pertmap}) has been performed, with $w_q=w_p=1$
and nonlinearity parameters $K_q$, $K_p$ randomly distributed between $-1/2$
and $+1/2$.  We note the expected exponential behavior for small $T_{\rm max}$,
with the classical decay rate $M/T_H$, as well as the rapid power-law decay of
the lost probability for $T_{\rm max}>T_H$, especially in the case of multiple
wave packets $M>1$.

\begin{figure}[ht]
\centerline{\includegraphics[width=4.5in,angle=270]{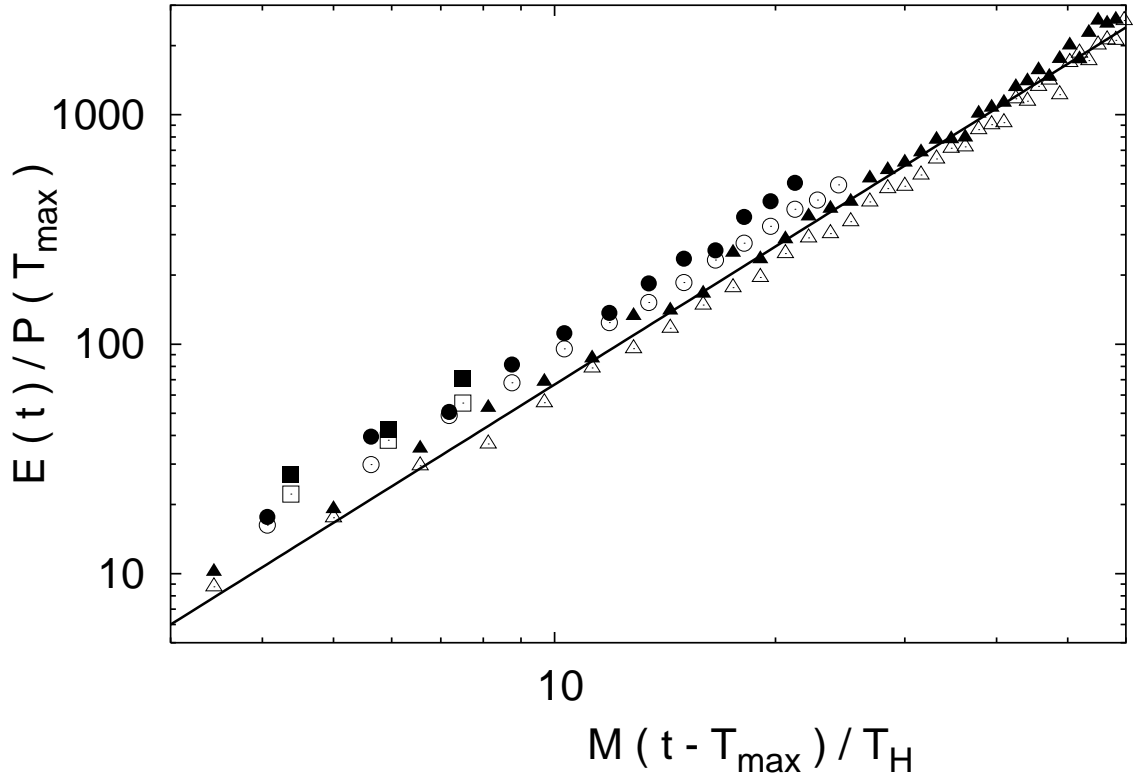}}
\vskip 0.2in
\caption{The ratio of the relative error in the propagator, $E(t)$, to the
probability lost on average during each cycle of recurrences, $P(T_{\rm max})$,
is plotted for times $t > T_{\rm max}$, $M=2$, several values of $T_{\rm max}$,
and two system sizes.  All system parameters are the same as in the previous
figure.  Squares, circles, and triangles correspond to $M T_{\rm max}/T_H=5$,
$10$, and $20$, respectively, while open and closed symbols distinguish system
size $N=64$ from system size $N=128$.  The solid line indicates the quadratic
growth of the error consistent with Eq.~(\ref{relerr}).
}
\label{fig_timerr}
\end{figure}

We are interested in the error induced at long times $t \gg T_{\rm max}$ by
omitting new recurrences not captured in the short-time correlation function.
At this point, we have not introduced any smoothing of the input data, i.e.,
$\tau=\infty$ in Eq.~(\ref{bltau}).  The typical returning amplitude at time
$t$ has completed $O(Mt/T_H)$ cycles of leaving and returning to the subspace
spanned by the $M$ wave packets $\phi_i$, i.e., in Eq.~(\ref{csum}) the
dominant terms are ones involving a product of $O(Mt/T_H)$ ${\mathbf
B}-$matrices.  In each cycle, probability given by Eq.~(\ref{plost}) is lost,
with the errors accumulating coherently, so that the relative error in the
matrix elements at time $t \gg T_{\rm max}=L T_0 \ge T_H/M$ is given by
\begin{equation}
E(t)={||{\mathbf C}_{L,\infty}(t) -{\mathbf C}(t)||^2 \over ||{\mathbf
C}(t)||^2 } \sim \left ({Mt \over T_H}\right)^2 P(T_{\rm max}) \,,
\label{relerr}
\end{equation}
where $||{\mathbf C}(t)||^2 = {\rm Tr} \; {\mathbf C}(t)^\dagger {\mathbf C}(t)
= \sum_{ij} |C_{ij}(t)|^2$.  The quadratic growth in the long-time error is
clearly seen in Fig.~\ref{fig_timerr} for several choices of the bootstrapping
parameters.

\begin{figure}[ht]
\centerline{\includegraphics[width=4.5in,angle=270]{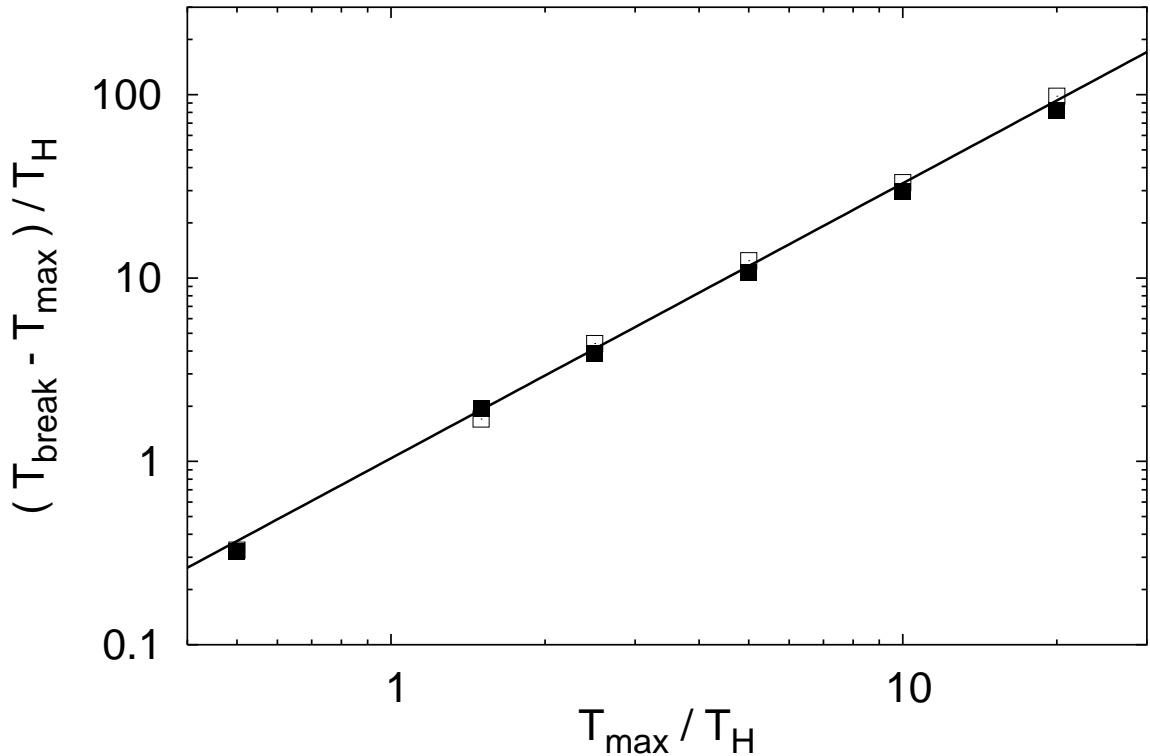}}
\vskip 0.2in
\caption{
The break time $T_{\rm break}$ of the bootstrapping approximation, defined by
$E(T_{\rm break})=1$, is plotted for $M=2$ and several choices of $T_{\rm
max}$.  Open and closed squares represent system sizes $N=64$ and $N=128$,
respectively.  The ensemble of systems is the same as in the previous two
figures.  The straight line is the theoretical prediction $T_{\rm break} \sim
T_{\rm max}^{(M+1)/2} \sim T_{\rm max}^{3/2}$ of Eq.~(\ref{tbreakpower}).
}
\label{fig_tbreak}
\end{figure}

To find the time scale $T_{\rm break}$ beyond which the bootstrapping procedure
breaks down, we assume $T_H/M  \le T_{\rm max} \le T_H/\sqrt{M}$.  Then,
setting the right hand side of Eq.~(\ref{relerr}) to unity and using
Eq.~(\ref{plost}), we obtain
\begin{equation}
T_{\rm break} \sim T_{\rm max} {\exp{({1 \over 2}M T_{\rm max}/T_H}) \over M
T_{\rm max} / T_H } \,.
\end{equation}
We see that including only a minimal number of new recurrences by setting
$T_{\rm max} \sim T_H/M$ leads to breakdown of the bootstrapping approximation
soon thereafter ($T_{\rm break} \sim T_{\rm max}$), but including additional
recurrences leads to exponential growth in the accuracy of the bootstrapping
approximation and consequently to exponential increase in the breakdown time.
Of course, this exponential growth ceases at very large values of $T_{\rm
max}$, when the error becomes dominated by a small fraction of eigenstates that
have unusually little overlap with the wave packets $\phi_i$.  Then $P(T_{\rm
max})$ follows the power-law behavior of Eq.~(\ref{plost}), and the growth in
$T_{\rm break}$ accordingly crosses over to a power-law behavior with $T_{\rm
max}$:
\begin{equation}
T_{\rm break} \sim T_H
{\sqrt{M+1} \over M} \left( {T_{\rm max} \over
T_H} \right ) ^{M+1 \over 2}
\label{tbreakpower}
\end{equation}
for $T_{\rm max} > T_H$.  We note that the growth of the break time $T_{\rm
break}$ with increasing bootstrap time $T_{\rm max}$ remains faster than linear
except in the single-wave packet case $M=1$.  This super-linear growth is
illustrated in Fig.~\ref{fig_tbreak} for the case of two wave packets ($M=2$),
where the break time $T_{\rm break}$ has been quantified as the time scale
where the relative error of Eq.~(\ref{relerr}) reaches unity.

\subsection{Results in the Energy Domain}
\label{secenergydom}

Starting with known short-time information about the correlation function, the
bootstrapped long-time dynamics may be Laplace or Fourier transformed into the
energy domain to obtain good approximations to the Green's function, spectrum,
or local density of states.  Alternatively, the short-time ``new" recurrences
may be transformed directly into the energy domain to obtain spectral
information, as indicated by Eqs.~(\ref{gtildeb}) and (\ref{tildeb}).  To avoid
unphysical oscillations in the spectrum on energy scales below $\hbar/T_{\rm
break}$ (associated with the breakdown of the bootstrapping approximation at
long times), we impose an explicit smooth cutoff on the the short-time
dynamics, in accordance with Eq.~(\ref{bltau}).  Loss of information is
minimized by choosing the cutoff time $\tau$ of the order of $T_{\rm break}$,
which is equivalent to Lorentzian smoothing of the spectrum on the scale
$\hbar/T_{\rm break}$.

\begin{figure}[ht]
\centerline{\includegraphics[width=4.5in,angle=270]{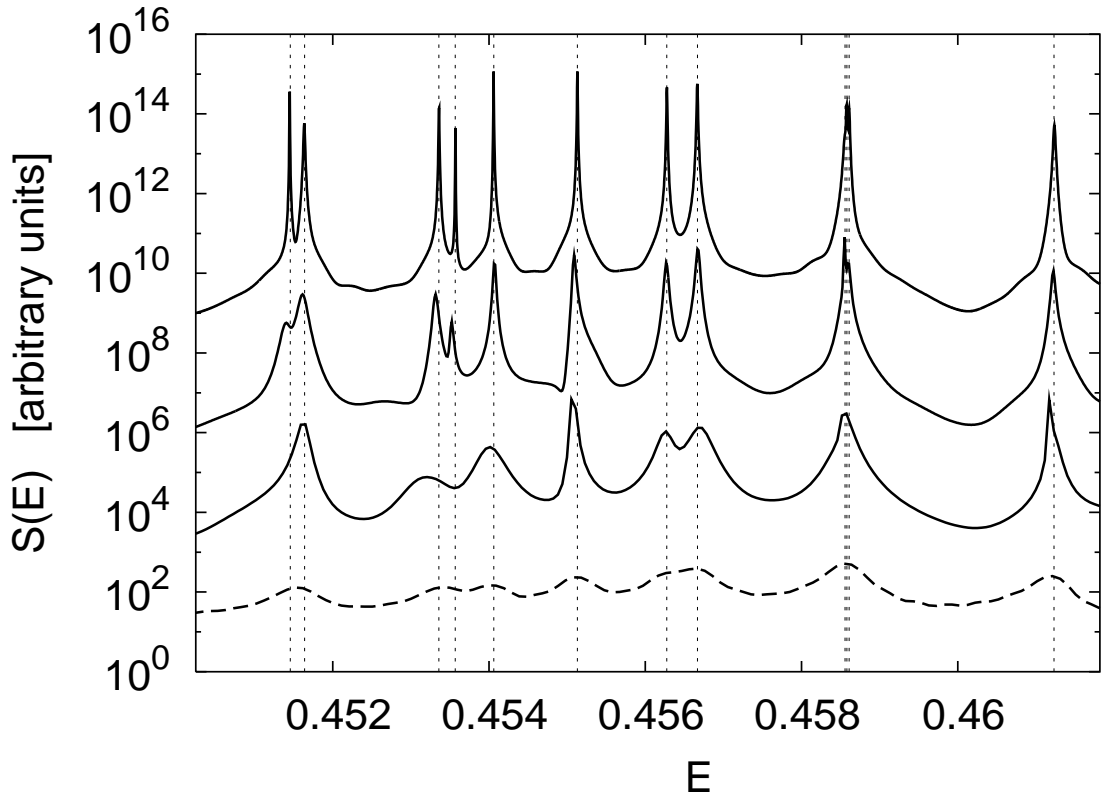}}
\vskip 0.2in
\caption{The local density of states summed over $M=6$ wave packets located at
energy $E=0.46$ in the Barbanis potential of Eq.~(\ref{barbeqn}),
$S(E)=\sum_{j=1}^6 {\rm Re}\;  (i/\pi) \, G_{jj}(E)$, is computed using the
bootstrapping approximation in accordance with Eq.~(\ref{gtildeb}).  The three
solid curves from bottom to top correspond to different bootstrap times used in
the bootstrapping calculation: $T_{\rm max}=T_H/2$, $T_{\rm max}=T_H$, and
$T_{\rm max}=2 T_H$.  For comparison, the result of Laplace transforming the
correlation function through time $T_{\rm max}=2 T_H$, without bootstrapping,
is shown as a dashed curve.  Each spectrum has been scaled by an arbitrary
constant to allow for easy comparison on a single plot.  In each case a
smoothing time scale $\tau \sim T_{\rm break}$ has been chosen to remove
unphysical oscillations in the spectrum in the energy range shown.  The dotted
vertical lines indicate the locations of the exact resonance peaks, obtained by
taking $T_{\rm max} \to \infty$.
}
\label{fig_spec}
\end{figure}

The numerical data presented in Fig.~\ref{fig_spec} is obtained for the
Barbanis potential, with parameters $a=1.1$ and $\hbar=0.0198$ in
Eq.~(\ref{barbeqn}), corresponding to slightly over $300$ quantum resonances in
the metastable well.  Six initial Gaussian wave packets are used in the
calculation, all centered at $x=0$ and having an average momentum of magnitude
$|p|=0.96$, corresponding to an energy $E=0.46$.  Thus, we are viewing dynamics
slightly below the top of the barrier, $E_{\rm max}=0.5$.  The classical
dynamics in the energy range considered is approximately $57\%$ chaotic, as
measured using a Poincar\'e surface of section at $x=0$.  All six initial wave
packets are centered in the chaotic region of classical phase space.

We see from the middle solid curve in Fig.~\ref{fig_spec} that most peaks in
the spectrum can readily be resolved using bootstrapping, taking correlation
information up through the Heisenberg time $T_H$ as our only input.  For
bootstrap time $T_{\rm max}=2 T_H$ (top solid curve), the spectral peak heights
already stand out by four orders of magnitude above the background.  The root
mean squared error in the bootstrap-predicted peak locations drops from $0.034
\Delta$ when $T_{\max}=T_H$ to $0.0064 \Delta$ for $T_{\rm max}=2T_H$, where
$\Delta$ is the mean level spacing.  We may contrast this with the result,
indicated by the dashed curve, of merely transforming and smoothing the same
correlation information, up through $T_{\rm max}=2T_H$, but without the benefit
of bootstrapping.  Here the resolution is very poor, and we are far from being
able to detect, for instance, the two doublets near $E=0.4515$ and $E=0.4535$.

In contrast with the map model studied in Section~\ref{secnoisefree}, in the
Hamiltonian system investigated here we must discretize time explicitly by
introducing a new time scale $T_0$.  The results of the calculation, however,
are unaffected by the choice of $T_0$, as long as $T_0<\hbar/\delta E$, where
$\delta E$ is the energy uncertainty in the wave packets $\phi_i$.
Equivalently, the time step $T_0$ must be chosen short enough so that the
self-overlaps $C_{ii}(T_0)$ are large due to free-flight dynamics.

A second key difference with the map model is that quantum motion in the
Barbanis potential is described by resonances rather than bound states.
Indeed, by comparing the upper two curves in Fig.~\ref{fig_spec}, we see that
in the $T_{\rm max}=T_H$ bootstrapped spectrum, the widths of several peaks
(e.g., the rightmost one near $E=0.4612$) are already dominated by the
intrinsic resonance widths rather than by any error associated with the time
cutoff.  In general, the efficiency of the bootstrapping approach increases as
one considers systems that are more open, since it is sufficient to choose a
bootstrap time $T_{\rm max}$ that will generate accurate dynamics to time
$T_{\rm break} \sim T_{\rm decay}$, where $T_{\rm decay}$ is the intrinsic
lifetime of the resonances, possibly shorter than $T_H$.

Finally, a third major difference between perturbed cat maps and the Barbanis
potential is the presence of regular as well as chaotic states in the Barbanis
spectrum.  By choosing the test wave packets $\phi_i$ appropriately, one may
optimally resolve states in the phase space region that are of greatest
interest in a given application.  For example, in ordinary scar theory, one may
begin with a wave packet centered on a specific periodic
orbit~\cite{scarhusimi}, with the aim of obtaining optimal information on the
structure of wave functions with high intensity on that orbit and their
associated eigenvalues; the price to be paid is the suppression of the
``anti-scarred" eigenstates that have anomalously low intensity on the same
orbit.  Here, we have randomly placed the six test wave packets in the chaotic
portion of phase space, improving our ability to resolve the chaotic states,
but necessitating the use of longer bootstrap times $T_{\rm max}$ to identify
spectral peaks associated with regular wave functions, such as the very narrow
resonance peak near $E=0.4536$.

In this context we note also that, in the case of scar theory, little or no
benefit is gained by following several wave packets launched along the same
weakly unstable periodic orbit, since they all exhibit very similar time
evolution, and share nearly identical local density of
states~\cite{scarmometer}.  From a bootstrapping perspective, we may consider
two wave packets nearly related by time evolution, e.g., $|\phi_2\rangle
\approx e^{-i\hat H \tau/\hbar}|\phi_1\rangle$.  Then the probability
$1-P(T_{\rm max})$ of Eq.~(\ref{probnoreturn}) for returning to the subspace
spanned by $\phi_1$ and $\phi_2$ by time $T_{\rm max}$ is nearly the same as
the probability of returning to $\phi_1$ itself, assuming $T_{\rm max} \gg
\tau$.  The rapid decrease in the ``lost probability" $P(T_{\rm max})$ with
increasing number of wave packets $M$, as indicated by Eq.~(\ref{plost}),
depends entirely on the wave packets behaving in an uncorrelated manner.  Thus,
the bootstrapping procedure for multiple wave packets is most effective when
the wave packets are chosen from different regions of phase space to avoid
obvious correlations.

\subsection{Influence of Noise}

\begin{figure}[ht]
\centerline{\includegraphics[width=4.5in,angle=270]{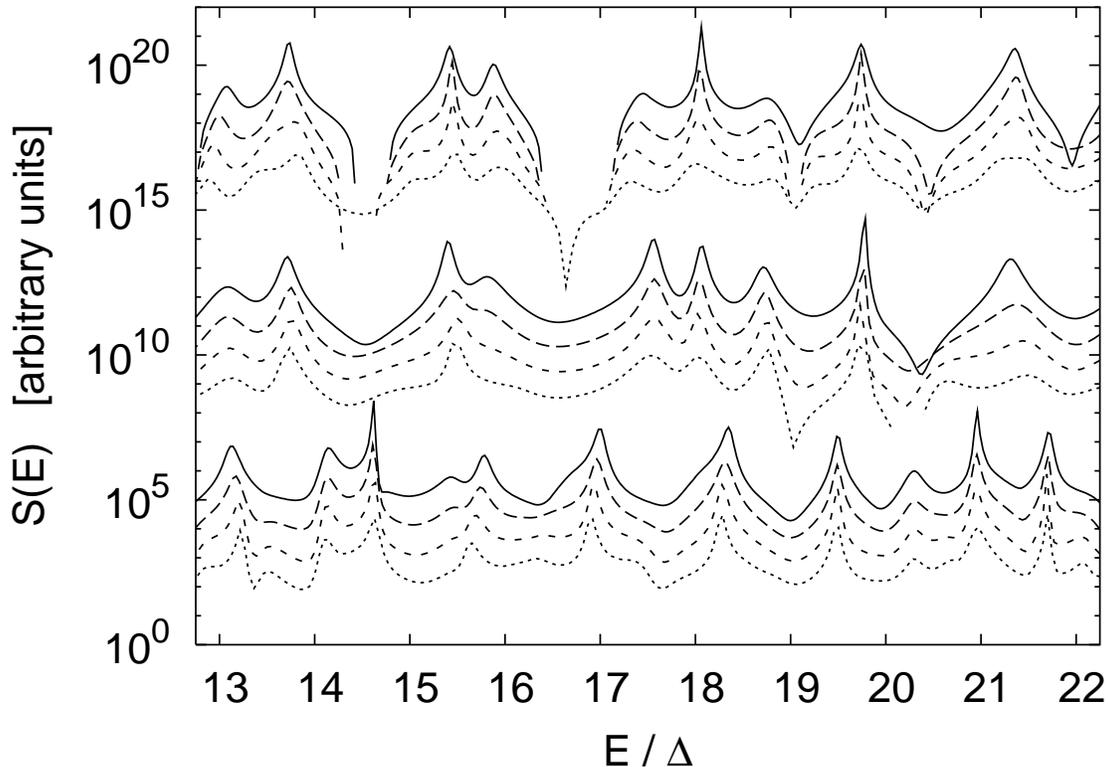}}
\vskip 0.2in
\caption{
The local density of states summed over $M$ randomly placed wave packets,
$S(E)= \sum_{j=1}^M {\rm Re} \; (i /\pi) \, G_{jj}(E)$, is computed for the
quantum map of Eq.~(\ref{pertmap}), with $w_q=w_p=1$, $K_q=0.2$, and
$K_p=-0.3$, using the bootstrapping approximation after noise has been
introduced into the short-time input data.  From top to bottom, the three sets
of curves correspond to (a) $N=128$, $M=2$, (b) $N=128$, $M=6$, and (c)
$N=512$, $M=2$, where $N$ is the system size or Hilbert space dimension.
Within each set, the top (solid) curve is the reconstructed spectrum in the
absence of noise, and the three dashed/dotted curves, from upper to lower,
indicate reconstructed spectra for the same system with the dimensionless noise
parameter set to $\epsilon=0.1$, $\epsilon=0.2$, and $\epsilon=0.3$.  Each
spectrum has been scaled by an arbitrary constant to allow for easy comparison
on a single plot.  In all cases, the spectrum is reconstructed from the
correlation function for $t\le T_{\rm max}=3 T_H/M$.
}
\label{fig_specerr}
\end{figure}

Noise in the input signal may be an important factor in specific applications
of the bootstrapping procedure, for example where a semiclassical or other
approximation is used to calculate the short-time correlation function.  We
return to the quantum map model discussed in Sec.~\ref{secnoisefree} and
introduce random noise into the short-time correlation function, as indicated
in Eq.~(\ref{defeps}).  The random error matrix elements $D_{ij}(m)$ in
Eq.~(\ref{defeps}) are independent Gaussian random variables of zero mean and
variance $1/N$, where $N$ is the Hilbert space dimension, so that
$\overline{|D_{ij}(m)|^2}=\overline{|C_{ij}(m)|^2}$ at long times $m$.  Then
the dimensionless parameter $\epsilon$ characterizes the relative size of the
noise.  A spectrum may be produced by bootstrapping the noisy short-time
data.  The results of such a calculation are presented in
Fig.~\ref{fig_specerr}.  We see that the spectral reconstruction is quite
robust for small $\epsilon$, and breaks down at around $\epsilon=0.2$
or $0.3$, independent of $N$ and $M$.

\begin{figure}[ht]
\centerline{\includegraphics[width=4.5in,angle=270]{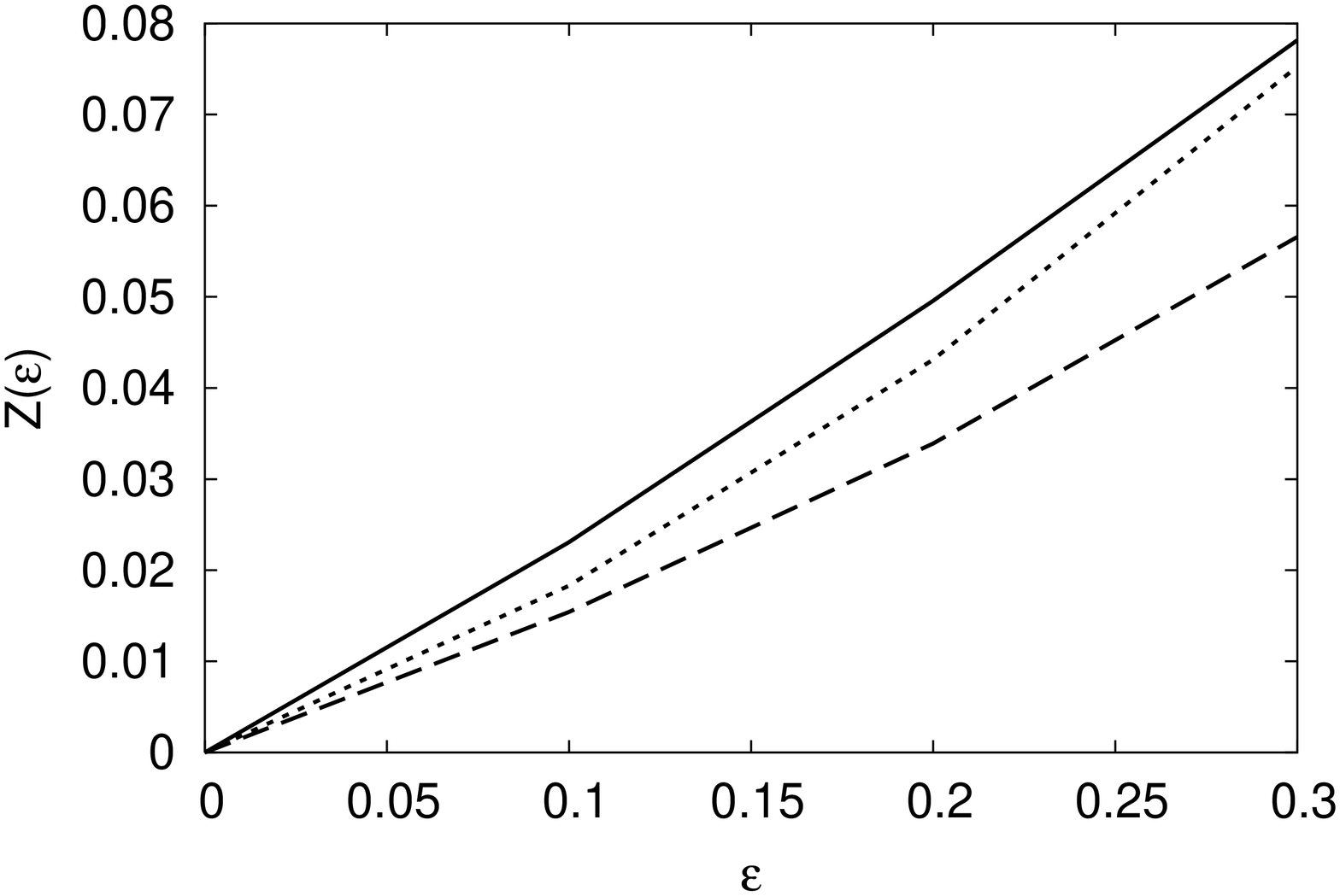}}
\vskip 0.2in
\caption{
The error $Z(\epsilon)$ in the reconstructed spectrum for the quantum map of
Eq.~(\ref{pertmap}) is computed as a function of the noise level $\epsilon$ in
the input correlation function.  All parameters are identical to those in
Fig.~\ref{fig_specerr}.  The solid curve indicates system size $N=128$ with
$M=2$ wave packets, the dashed curve is for $N=512$ with $M=2$, and the dotted
curve is for $N=128$ with $M=6$.  In all cases, the noise-free spectrum $S(E)$
and the noisy spectrum $S_\epsilon(E)$ are both reconstructed from the
correlation function for $t\le T_{\rm max}=3 T_H/M$.
}
\label{fig_specdata}
\end{figure}

To study more carefully the breakdown in the accuracy of the bootstrapped
spectrum and its dependence on parameters $N$ and $M$, we need to define
a quantitative measure of the error in the bootstrapped spectrum.  Consider a
local density of states $S(E)$ reconstructed from the exact correlation
function known for $t \le T_{\rm max}$ and a local density of states
$S_\epsilon(E)$ reconstructed from the same input but with added noise
characterized by $\epsilon$ as in Eq.~(\ref{defeps}).  We may define
the dimensionless error ratio
\begin{equation}
Z(\epsilon)={\int dE \,(\ln S_\epsilon(E)-\ln S(E))^2 \over
\int dE \, (\ln S(E))^2 } \,,
\end{equation}
which measures the error induced in the reconstructed spectrum by noise of size
$\epsilon$ in the input.  We note that it is appropriate to focus on the
logarithm of the reconstructed spectrum, because the spectrum itself is
dominated by sharp peaks, as seen in Fig.~\ref{fig_specerr}.  The quantity
$Z(\epsilon)$ is shown in Fig.~\ref{fig_specdata}, for the same parameters as
were used earlier in Fig.~\ref{fig_specerr}.  Not surprisingly, we observe
growth in the spectral error with increasing noise $\epsilon$, but, more
importantly, this error is almost independent of system size $N$ and number of
wave packets $M$ (at $\epsilon=0.3$, $Z(\epsilon)$ varies at most by $30\%$ as
$N$ changes by a factor of $4$ and $M$ by a factor of $3$).  The same results
have been observed for other system parameters.  This robustness implies that
input with noise of a small but finite size $\epsilon$ may be used in the
semiclassical limit $N \to \infty$ (equivalently, $\hbar \to 0$).  Obviously,
an even more favorable situation exists when the noise level $\epsilon$
decreases with increasing $N$.  An important example of the latter situation
exists when the ``noise" results from using a semiclassical (stationary phase)
approximation for the short-time dynamics $t \le T_{\rm max}$.  Then $\epsilon
\sim \hbar$, which decreases with $N$.  Therefore short-time
evolution computed within a semiclassical approximation may be used with
confidence to obtain stationary properties of the exact quantum system.

\section{Bootstrapped Wave Function Statistics}
\label{secstat}

We now consider the situation where known short-time dynamical information is
insufficient for resolving individual eigenstates, and the focus therefore
shifts to predictions of a statistical nature.  In other words, we consider an
ensemble of systems that all share (perhaps approximately) a given short-time
dynamics, and ask what information can be extracted about the distribution of
wave functions in systems drawn from this ensemble.

\subsection{Inverse Participation Ratio Calculations}

The simplest quantitative measure of wave function structure is the inverse
participation ratio (IPR) or second moment of the wave function
intensities~\cite{ipr}: ${\rm IPR}_\Psi= N \sum_{i=1}^N |\langle
\phi_i|\Psi\rangle|^4$, where $\Psi$ is an eigenstate, $N$ is the Hilbert space
dimension, the sum is over a complete basis $\phi_i$, and we impose the usual
normalization condition $\sum_{i=1}^N |\langle \phi_i |\Psi| \rangle|^2=1$.
The IPR measures the degree of wave function localization, ranging from $1$ for
a delocalized wave function having uniform overlaps with all basis states to
$N$ for a completely localized state $\Psi$.  RMT predicts ${\rm IPR}=2$ in the
absence of time-reversal or other symmetry.  Similarly, for each basis state
$\phi$ we may define a local IPR (LIPR) as ${\rm LIPR}_\phi = N\sum_{j=1}^N
|\langle \phi|\Psi_j\rangle|^4$, where the sum extends over
eigenstates~\cite{scarmom}; the LIPR measures the degree of localization
associated with a specific basis element $\phi$ and is proportional to the
average long-time return probability $|\langle \phi|\phi(t)\rangle|^2$ as $t
\to \infty$.

Extending arguments developed originally for periodic orbit
scars~\cite{scarhusimi,scarmom}, we may interpret long-time dynamics in a
chaotic system as a convolution of known short-time recurrences with
quasi-random long-time recurrences, 
\begin{equation}
\langle \phi|\phi(t)\rangle \approx  \sum_{\tau=-T}^{T} \langle
\phi|\phi(\tau)\rangle r_\phi(t-\tau) \,,
\label{timeconvol}
\end{equation}
where for simplicity we have assumed discrete-time dynamics, the sum over
$\tau$ extends to some scale $T$ that includes as much as possible of the
non-random dynamics of interest but is still short compared with the Heisenberg
time $T_H$, and $r_\phi(t')$ are Gaussian random independent variables,
associated with nonlinear long-time recurrences.  For the LIPR, we obtain
\begin{eqnarray}
{\rm LIPR}_\phi & \approx  & 2 \sum_{\tau=-T}^{T}
 |\langle \phi|\phi(\tau)\rangle|^2 \label{liprpred1} \\
& =&  2\left [1+2 \sum_{\tau=1}^{T} |C_{\phi \phi} (\tau)|^2\right]
\label{liprpred}
\end{eqnarray}
in the notation of Sec.~\ref{secboot}, where the overall prefactor of $2$ is
the RMT result in the absence of time-reversal symmetry.  The autocorrelation
function or return amplitude
$C_{\phi \phi}(\tau)$ may be computed from the ${\mathbf B}$-matrices
using the bootstrapping formulas of Eqs.~(\ref{bijdef}) and (\ref{cindex}), or we
may explicitly write
\begin{equation}
{\rm LIPR}_\phi \approx 2\left [1+2 \sum_{\tau=1}^{\infty}
\left |\sum_{\tau_1=1}^{T_{\rm max}} B_{\phi \phi}(\tau_1) \delta(\tau_1-\tau)+
\sum_{\tau_1,\tau_2=1}^{T_{\rm max}} \sum_{\phi'} B_{\phi \phi'}(\tau_2) B_{\phi'
\phi}(\tau_1)\delta(\tau_1+\tau_2-\tau)+ \cdots\right |^2 \right] \,.
\end{equation}
Here the upper limit $T$ in the sum over $\tau$ may safely be taken to
infinity, as long as the bootstrap time $T_{\rm max} \ll T_H/M$, so that
most of the probability is lost by the Heisenberg time, and times $\tau \sim T_H$
do not contribute significantly to the sum.  We note that
the second- and higher-order bootstrapping terms implicitly include revivals at
times longer than $T_{\rm max}$, although only the correlation function up to
$T_{\rm max}$ is used as input to the calculation.  The bootstrapping formula
makes optimal use of the available short-time information, and good agreement
may be obtained even for fairly short bootstrap times $T_{\rm max}$.

\begin{figure}[ht]
\centerline{\includegraphics[width=4.5in,angle=270]{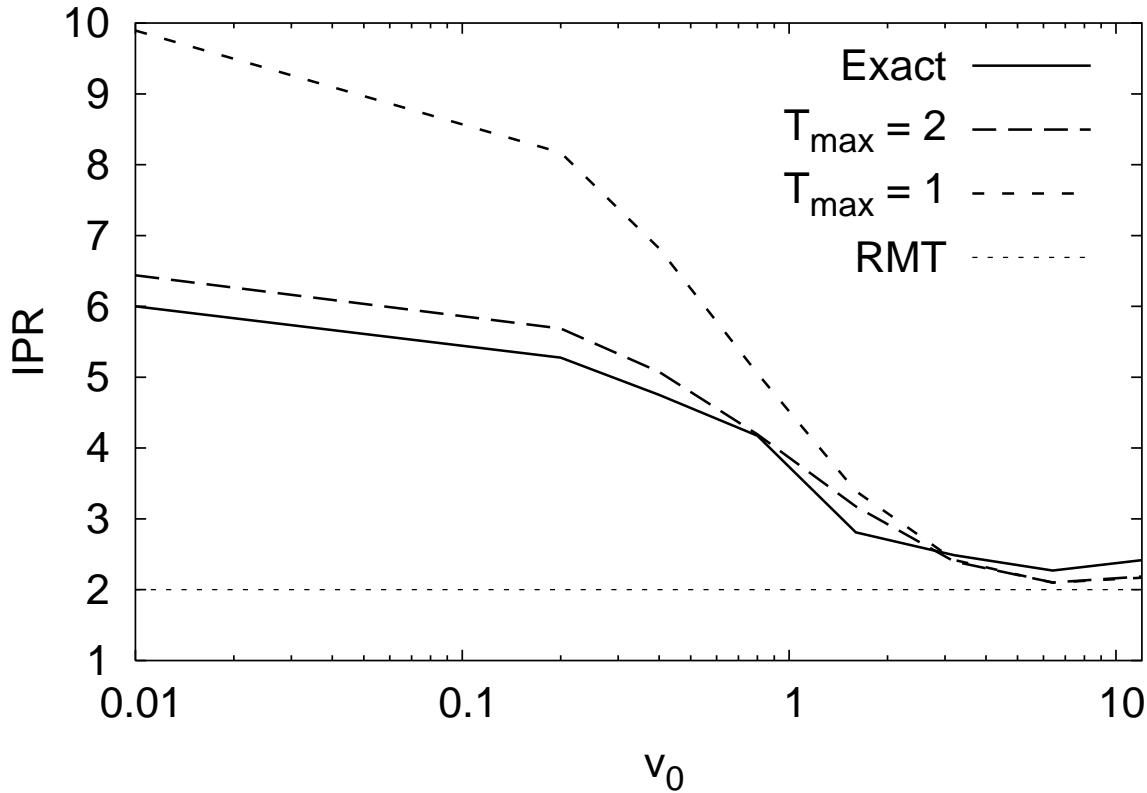}}
\vskip 0.2in
\caption{
The local inverse participation ratio (LIPR) at the location of the cusp is
calculated for the diffractive potential of Eq.~(\ref{diffrpot}), for system
size $N=64$, one wave packet centered on the cusp at $q_0=-0.2 \pi$, and
several values of the kick parameter $v_0$.  The exact data is averaged over
boundary conditions for each value of $v_0$.  The bootstrapping prediction,
using Eq.~(\ref{liprpred}) for one wave packet centered on the cusp, is shown
for $T_{\rm max}=1$ and $2$, in units of the kick period.  The RMT prediction,
equivalent to bootstrapping with $T_{\rm max}=0$, provides a baseline for
comparison.  }
\label{fig_ipr}
\end{figure}

As a specific example, we consider a quantum map defined by
Eq.~(\ref{quanmap}), with kinetic term $T(p)={1 \over 2}p^2$ and kicked
potential 
\begin{equation}
V(q)=-{1 \over 2} (q-q_0)^2+v_0 \left[{q \over q_0} \Theta(q-q_0)+
{1-q \over 1-q_0} \Theta(q_0-q)\right] \,,
\label{diffrpot}
\end{equation}
where as before $q$ and $p$ both range from $-\pi$ to $+\pi$, and $\Theta(x)$
is the usual step function defined by $\Theta(x)=1$ for $x >0$ and
$\Theta(x)=0$ otehrwise.  The dynamics is fully chaotic and has no period-one
classical orbits, but does have a diffractive orbit at $q=q_0$, $p=0$,
associated with a cusp in the kick potential.  The bootstrapping calculation is
performed for a single wave packet centered on this diffractive orbit.  In
Fig.~\ref{fig_ipr}, we calculate the LIPR for this wave packet, as a function
of parameter $v_0$, exactly and in the bootstrapping approximation.  We see
immediately that RMT (equivalent to the bootstrapping prediction with $T_{\rm
max}=0$) severely underestimates the degree of wave function localization when
$v_0 < 1$ and the diffractive orbit is consequently strong.  Bootstrapping the
one-step recurrence only ($T_{\rm max}=1$) greatly overestimates the effect,
but the $T_{\rm max}=2$ calculation, which incorporates information about
one-step and two-step new recurrences, already gives a good approximation to
the exact answer over the entire range of $v_0$.  We note that the
bootstrapping has been performed here using one- and two-step time correlation
data for a single wave packet; obviously the results can only improve if
multiple wave packets are used simultaneously with the same bootstrap time
$T_{\rm max}$.

\begin{figure}[h]
\centerline{\includegraphics[width=4.5in,angle=270]{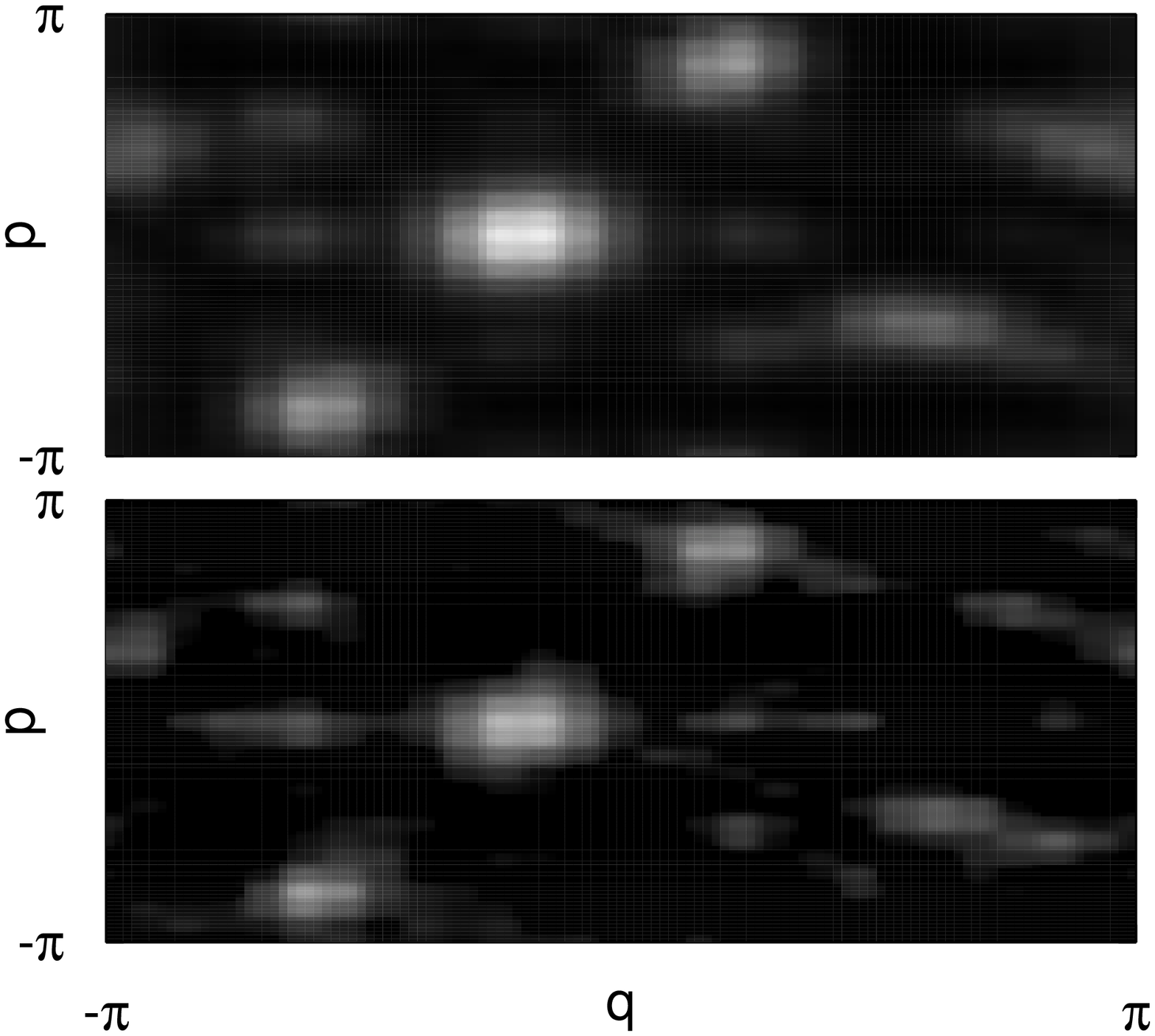}}
\vskip 0.2in
\newpage
\caption{
The local inverse participation ratio (LIPR) is plotted as a function of
position and momentum for the potential of Eq.~(\ref{diffrpot}), with system
size $N=64$, cusp location $q_0=-0.2\pi$, and cusp strength $v_0=0.59$.  The
exact LIPR landscape is shown in the upper panel, while the lower panel
represents the prediction of a bootstrapping procedure with $T_{\rm max}=3$.
The color scale ranges from LIPR=2 (black) to LIPR=5 (white).
}
\label{fig_iprgrid}
\end{figure}

We now fix $v_0=0.59$ and repeat the above bootstrapping calculation for single
wave packets centered at various locations in phase space.  In each case, we
find the exact LIPR by diagonalizing the evolution matrix.  We also predict the
LIPR using the bootstrapping approximation with $T_{\rm max}=3$, i.e., the
recurrences for three time steps are computed exactly, bootstrapped to obtain
long-time behavior, and then used to estimate the local inverse participation
ratio in accordance with Eq.~(\ref{liprpred}).  The results are shown in
Fig.~\ref{fig_iprgrid}.  Here, the bright spot slightly to the left of center is
the localization peak associated with a diffractive orbit at $q=q_0=-0.2\pi$,
$p=0$.  We observe that the bootstrapping procedure allows not only this peak
but most significant features of the localization landscape to be well resolved
by $T_{\rm max}=3$.

\subsection{Wave Function Intensity Distribution}

A prescription similar to the above may be used to compute higher moments of
the intensity distribution beyond the standard inverse participation ratio;
instead, we turn our attention to the intensity distribution itself.  Knowledge
of such a distribution is essential, for example, to the understanding of
resonance width or decay rate statistics in a weakly open system.  In the
context of scarring, it has been shown that the probability distribution of
wave function intensities may be obtained by combining a smooth spectral
envelope constructed from the short time dynamics with Gaussian random
fluctuations on fine energy scales~\cite{scarinten}.  More generally, whenever
a separation of scales exists between non-random short-time dynamics and
quasi-random long-time behavior, we may write the local density of states
(strength function) for wave packet $\phi$ as~\cite{papenbrock} 
\begin{equation}
{\rm Re} \;(i / \pi)\,G_{\phi\phi}(E)=\sum_n \delta(E-E_n) |\langle
\phi|\Psi_n\rangle|^2 \approx \sum_n \delta(E-E_n) S^{\rm smooth}_\phi(E)
|R_n|^2 \,,
\label{specmult}
\end{equation}
where
\begin{equation}
S^{\rm smooth}_\phi(E)=\sum_{\tau=-T}^T \langle \phi|\phi(t)\rangle e^{iEt/\hbar}
\end{equation}
is a Fourier transform of the short-time signal and $R_n$ are independent
Gaussian random variables with variance $1/N$ (real or complex depending on the
presence or absence of time reversal symmetry, respectively).  The above
expressions assume no symmetry, with the possible exception of time reversal,
and must be appropriately modified in the presence of such symmetries,
including parity invariance~\cite{scarmom}.  The multiplication in
Eq.~(\ref{specmult}) of a known short-time signal by a long-time signal assumed
to be quasi-random is the energy-domain counterpart of the convolution formula
appearing in Eq.~(\ref{timeconvol}).

In the bootstrapping context, we may obtain the short-time envelope using
Eqs.~(\ref{gtildeb}) and (\ref{tildeb}), where exact new recurrences ${\mathbf
B}(m)$ are replaced by ${\mathbf B}_{L,\tau}(m)$ as defined by
Eq.~(\ref{bltau}) for some choice of $T_{\rm max}=LT_0$ and a smoothing time
scale $\tau$.  This is the same procedure we used to construct approximate
bootstrapped spectra in Sec.~\ref{secenergydom}, except that there the
bootstrap time $T_{\rm max}$ was chosen sufficiently long to resolve individual
states, $T_{\rm max} > T_H/M$, while here we may take $T_{\rm max}$ to be only
a small multiple of the one-step time $T_0$.

Once a short-time local density of states envelope $S^{\rm smooth}_\phi(E)$ is
known, we may directly construct the probability distribution of wave function
intensities $I=|\langle \phi|\Psi_n\rangle|^2$.  We need only to multiply the
envelope heights $S^{\rm smooth}_\phi(E)$, with uniformly distributed energies
$E$, by random factors $|R|^2$ where $R$ is Gaussian-distributed (and $|R|^2$
is therefore exponentially distributed for complex $R$):
\begin{equation}
P(I)={1 \over {E_2-E_1} } \int_{E_1}^{E_2} dE \int_0^\infty d(|R|^2) e^{-|R|^2}
\delta \left(I-S^{\rm smooth}_\phi(E)|R|^2 \right ) \,.
\label{bootpi}
\end{equation}

\begin{figure}[ht]
\centerline{\includegraphics[width=4.5in,angle=270]{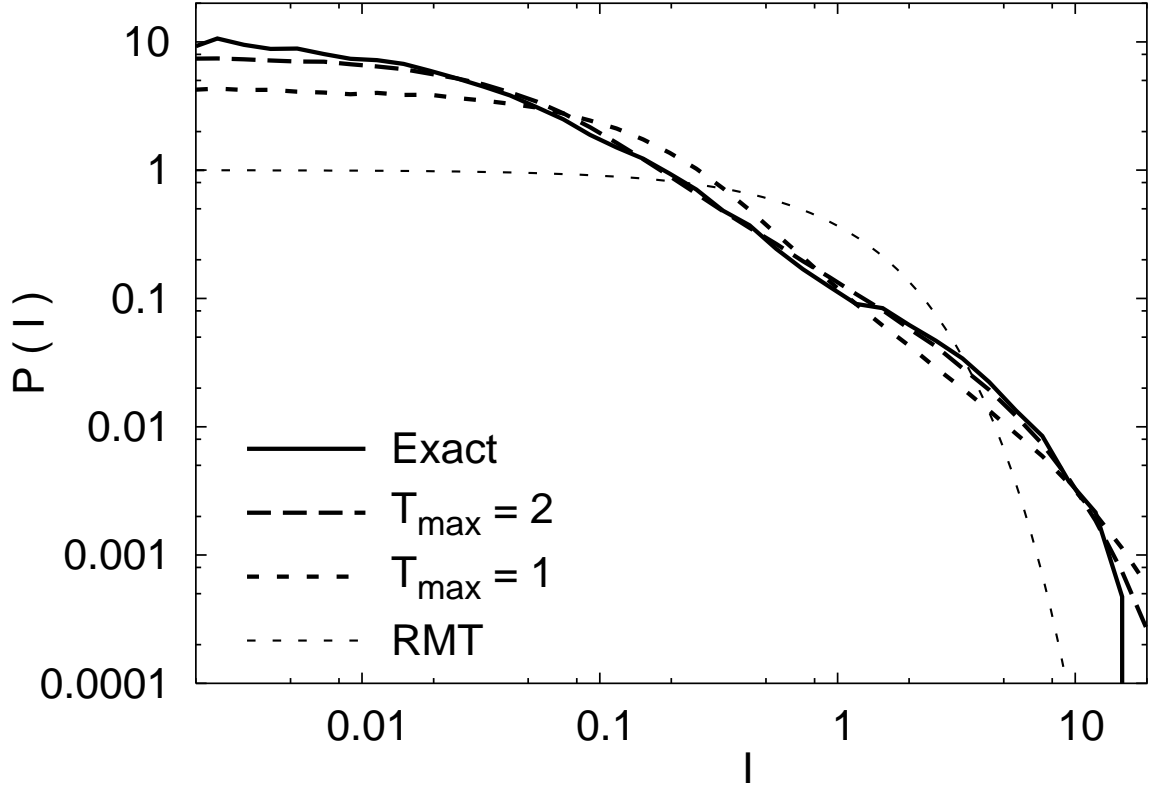}}
\vskip 0.2in
\caption{
The distribution of wave function intensities $I=|\langle
\phi|\Psi_n\rangle|^2$ for the kicked map of Eq.~(\ref{diffrpot}) with system
size $N=64$, cusp location $q_0=-0.2\pi$, and kick parameter $v_0=0.2$ is
shown, where $\phi$ is a Gaussian wave packet centered on the cusp at $q=q_0$,
$p=0$.  The solid curve shows the exact data, obtained by diagonalizing time
evolution matrices, and averaging over systems with different boundary
conditions.  The bootstrapped predictions are computed using
Eq.~(\ref{bootpi}), where the smooth envelope is obtained from bootstrap time
$T_{\rm max}=1$ or $T_{\rm max}=2$, for a single wave packet $M=1$.  The RMT
prediction, equivalent to bootstrapping with $T_{\rm max}=0$, provides a
baseline for comparison.  }
\label{fig_inten}
\end{figure}

Typical examples of the resulting intensity distribution are shown in
Fig.~\ref{fig_inten}.  Here we use the same system and wave packet location as
in Fig.~\ref{fig_ipr}, but fix kick parameter $v_0$ at the value $0.2$.  The
short-time envelope $S^{\rm smooth}_\phi$ is constructed either using only
one-step new recurrences (bootstrap time $T_{\rm max}=T_0=1$) or using one- and
two-step new recurrences (bootstrap time $T_{\rm max}=2T_0=2$).  The $T_{\rm
max}=1$ short-time envelope already results in a predicted intensity
distribution that is a great improvement over the RMT prediction, correctly
predicting an excess of very large and very small wave function intensities at
the cusp.  The $T_{\rm max}=2$ envelope predicts an intensity distribution that
is in even better agreement with actual data.

\subsection{Wave Function Correlations}

The bootstrapping approach lends itself naturally to the consideration of
observables beyond the statistics of individual wave function intensities
$I=|\langle \phi|\Psi \rangle|^2$.  As a simple example, we may consider the
covariance ${\rm COV}_{\phi_1,\phi_2}=N\sum_{j=1}^N |\langle \phi_1|\Psi_j
\rangle|^2 |\langle \phi_2|\Psi_j \rangle|^2-1$, where $\phi_1$ and $\phi_2$
are two wave packets and the sum is once again over the eigenstates.  Obviously
the covariance is a generalization to two wave packets of the local inverse
participation ratio discussed earlier: ${\rm LIPR}_\phi= {\rm
COV}_{\phi,\phi}+1$.  The covariance or correlation between wave function
intensities at two points is clearly important, for example, for understanding
the statistics of conductance peak heights in a weakly open quantum dot with
two leads~\cite{scarcond}; it is also relevant for analogous reaction rate
calculations or for the computation of interaction matrix elements.

Letting $|\phi\rangle={1 \over \sqrt{2}}
(|\phi_1\rangle+e^{i\theta}|\phi_2\rangle)$, using Eq.~(\ref{liprpred1}) for
${\rm LIPR}_\phi$, and averaging over the relative phase $\theta$, we obtain
\begin{eqnarray}
{\rm COV}_{\phi_1,\phi_2} & \approx  & \sum_{\tau=-T}^{T}
 \left [|\langle \phi_1|\phi_2(\tau)\rangle|^2
+ \langle \phi_1|\phi_1(t)\rangle \langle \phi_2(t) |\phi_2\rangle \right ] -1
  \nonumber\\
& =&  |C_{\phi_1 \phi_2}(0)|^2+ 2 \sum_{\tau=1}^{T} |C_{\phi_1 \phi_2}
(\tau)|^2 + 2 \;{\rm Re} \sum_{\tau=1}^T C_{\phi_1 \phi_1}(\tau) C^\ast_{\phi_2 
 \phi_2}(\tau) \label{covpred}
\end{eqnarray}

Two types of terms are present in Eq.~(\ref{covpred}): ones associated with the
short-time probability for evolving from state $\phi_1$ to state $\phi_2$ or
vice versa, and ones associated with a correlation between the individual
short-time autocorrelation functions for $\phi_1$ and $\phi_2$.  Once again,
the correlation functions $C_{\phi \phi'}(\tau)$ may be computed using the
bootstrapping formula given by Eq.~(\ref{cindex}) or Eq.~(\ref{csum}), where
the ``new" recurrences ${\mathbf B}(m)$ are known up to the bootstrap time
$T_{\rm max}=LT_0$, as in Eq.~(\ref{bltau}).  As in the LIPR calculation, the
upper limit $T$ of the sum in Eq.~(\ref{covpred}) may be taken to infinity, as
long as $T_{\rm max} <T_H/M$.  For the covariance calculation, it is most
convenient to perform the bootstrapping with just $M=2$ initial wave packets:
$\phi_1$ and $\phi_2$.

\begin{figure}[ht]
\centerline{\includegraphics[width=4.5in,angle=270]{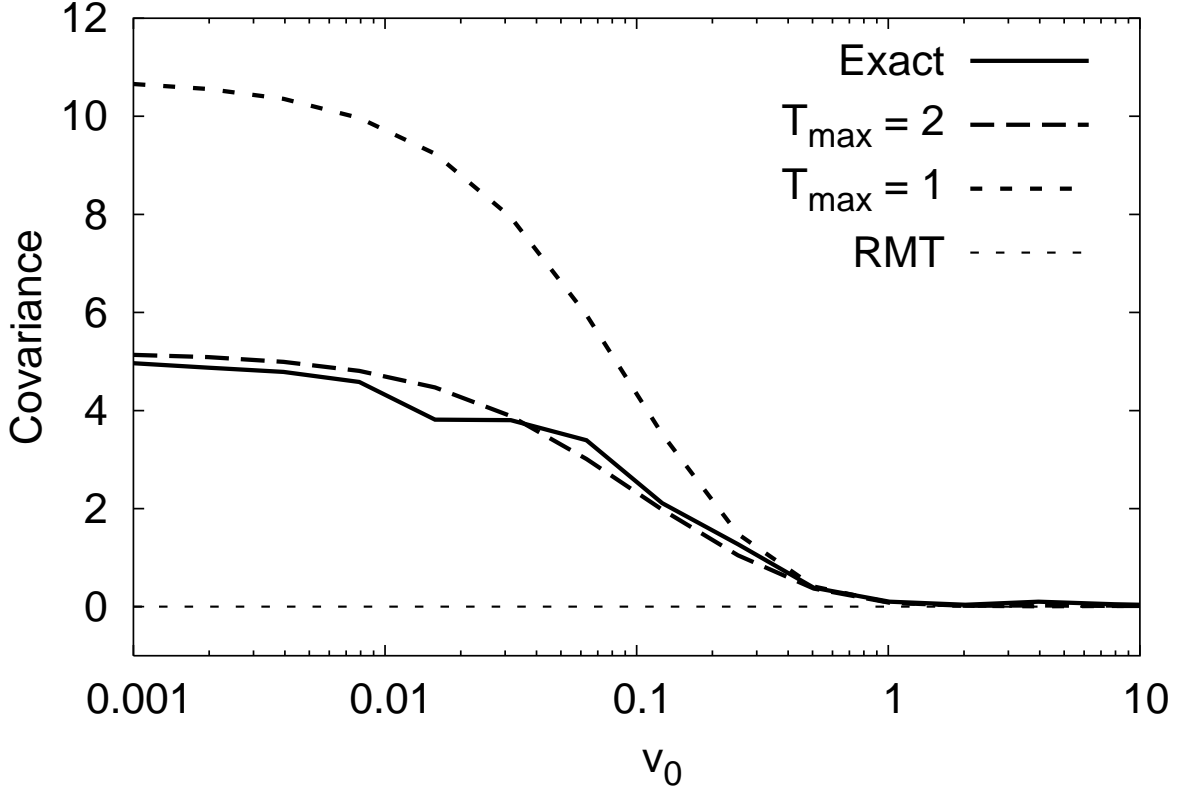}}
\vskip 0.2in
\caption{The covariance between wave function intensities $|\langle \phi_1
|\Psi\rangle|^2$ and $|\langle \phi_2 |\Psi\rangle|^2$, where $\phi_1$ and
$\phi_2$ are Gaussian wave packets centered at $(q=\pm \pi/2,p=p_0)$, is
computed for a quantum map with kinetic term given by Eq.~(\ref{twocuspkin})
and kick potential given by Eq.~(\ref{twocusp}).  Once again, the data is
averaged over boundary conditions for each value of the kick parameter $v_0$.
The system size is fixed at $N=256$, and the classical system parameters are
$a=1/4$, $b=1/20$, $q_0=-\pi/5$ and $p_0=-3 \pi/5$.  The exact results of
matrix diagonalization are compared with the bootstrapping calculation for
$M=2$ wave packets with bootstrap time $T_{\rm max}=1$ or $2$, in units of the
kick period. 
}
\label{fig_covar}
\end{figure}

As an example, we consider another quantum map, defined by Eq.~(\ref{quanmap})
with kinetic term
\begin{equation} 
T(p)={1 \over 2} (p-p_0)^2+ b \left [ \sin {2 (p-p_0)} -{1
\over 2} \sin{4(p-p_0)} \right]
\label{twocuspkin}
\end{equation} and periodic kick
\begin{equation}
V(q)=\left \{
\begin{array}{ll}
-{a \over 2}(q+\pi/2)^2+v_0 {q+\pi \over \pi/2} & -\pi< q<-\pi/2 \\
-{a \over 2}(q+\pi/2)^2+v_0 {q_0-q \over q_0+\pi/2} & -\pi/2<q <q_0 \\
-{a \over 2}(q-\pi/2)^2+{3 \over 2}v_0 {q-q_0 \over \pi/2 -q_0} & q_0 < q< \pi/2 \\
-{a \over 2}(q-\pi/2)^2+{3 \over 2}v_0 { \pi-q \over \pi/2} & \pi/2 < q <\pi
\end{array} \right.
\label{twocusp}
\end{equation}
This potential has a cusp-like maximum of height $v_0$ at $q=-\pi/2$ and
another of height $3v_0/2$ at $q=\pi/2$, resulting in the possibility of
diffractive periodic motion between $(q=-\pi/2,p=p_0)$ and $(q=\pi/2,p=p_0)$.
We compute the covariance between wave function intensities $|\langle \phi_1
|\Psi\rangle|^2$ and $|\langle \phi_2 |\Psi\rangle|^2$, where $\phi_1$ and
$\phi_2$ are Gaussian wave packets centered at the two points in phase space.
The results are presented in Fig.~\ref{fig_covar}, as a function of the cusp
height parameter $v_0$.  Once again, the bootstrapping prediction is shown for
bootstrap time $T_{\rm max}=1$ or $2$, in units where the kick period $T_{\rm
kick}$ is set to unity.  The RMT prediction, corresponding to bootstrap time
$T_{\rm max}=0$, is shown for comparison.  Just as in the LIPR and intensity
distribution calculations, rapid convergence is observed with increasing
$T_{\rm max}$, and almost all relevant information is already obtained by
bootstrapping the one-step and two-step dynamics.

\section{Summary}
\label{secsum}

Short-time dynamical information, either of classical origin or otherwise,
inevitably leaves its imprint on the long-time behavior and stationary
properties of a quantum system.  The bootstrapping approach allows this
information to be processed systematically, for one or an arbitrary number $M$
of initial wave packets.  Because multiple iterations of the known short-time
dynamics are included, the resulting spectral accuracy can be much greater than
what one would obtain, for example, by a simple Fourier transform of a
short-time signal.  At the same time, the procedure is extremely efficient,
requiring at each energy linear algebra operations involving only $M$ by $M$
matrices, and independent of the total size $N$ of the Hilbert space.  There is
no assumption of unitarity in the dynamics, and the procedure works equally
well for closed or open systems.  Robustness to errors in the short-time signal
implies, for example, that reliable calculations can be performed when the
short-time correlations are computed in a small-$\hbar$ or other approximation
relevant to a given problem.

The bootstrap time $T_{\rm max}$ can be varied to extract maximum information
from the least amount of input data.  At small values of $T_{\rm max}$, the
approach can be viewed as a generalization of standard periodic orbit scar
theory, leading to statistical prediction beyond RMT for local density of
states and wave function statistics.  Reliable quantitative predictions can be
obtained for inverse participation ratios, wave function intensity
distributions, and wave function correlations, even when the short-time
dynamics is of non-classical origin.  Increasing either $T_{\rm max}$ or $M$
allows for a systematic inclusion of additional correlations.  Once the product
$M T_{\rm max}$ becomes comparable to the Heisenberg time $T_H$, it becomes
possible to go beyond statistical predictions to resolve individual
eigenstates and energy levels, with an accuracy scaling exponentially with $M
T_{\rm max}/T_H$.  The initial wave packets $\phi_i$ can be chosen optimally to
minimize redundancy in the short-time correlations, and to obtain maximal
information in a specific basis or for wave function structure in a given
subspace of the original Hilbert space.

\begin{acknowledgments} 
Very useful discussions with E. J. Heller and W. E. Bies are gratefully
acknowledged.
\end{acknowledgments}

\end{document}